\documentclass{llncs}
\usepackage{graphicx}
\begin{document}

\title{I know what you mean: semantic issues in Internet-scale
publish/subscribe systems}
\author{Ioana Burcea \and Milenko Petrovic \and Hans-Arno Jacobsen}
\institute{Department of Electrical and Computer Engineering\\
Department of Computer Science\\
University of Toronto, Canada\\
\email{\{ioana,petrovi,jacobsen\}@eecg.toronto.edu}}

\maketitle

\begin{abstract}

In recent years, the amount of information on the Internet has
increased exponentially developing great interest in selective
information dissemination systems. The publish/subscribe paradigm is
particularly suited for designing systems for routing information and
requests according to their content throughout wide-area network of
brokers.  Current publish/subscribe systems use limited syntax-based
content routing but since publishers and subscribers are anonymous and
decoupled in time, space and location, often over wide-area network
boundary, they do not necessarily speak the same language.
Consequently, adding semantics to current publish/subscribe systems is
important. In this paper we identify and examine the issues in
developing semantic-based content routing for publish/subscribe broker
networks.

\end{abstract}

\section{Introduction}

The increase in the amount of data on the Internet has led to the
development of a new generation of applications based on selective
information dissemination where data is distributed only to interested
clients. Such applications require a new middleware architecture that can
efficiently match user interests with available information. Middleware
that can satisfy this requirement include event-based architectures such
as publish/subscribe systems.

In publish/subscribe systems (hereafter referred to as pub/sub
systems), clients are autonomous components that exchange information
by publishing events and by subscribing to the classes of
events\footnote{We use the terms \textit{event} and
\textit{publication} interchangeable.} they are interested in. In these
systems, publishers produce information, while subscribers consume it.
A component usually generates a message when it wants the external
world to know that a certain event has occurred. All components that
have previously expressed their interest in receiving such events will
be notified about it. The central component of this architecture is the
event dispatcher (also known as event broker). This component records
all subscriptions in the system. When a certain event is published, the
event dispatcher matches it against all subscriptions in the system.
When the incoming event verifies a subscription, the event dispatcher
sends a notification to the corresponding subscriber.

The earliest pub/sub systems were subject-based. In these systems, each
message (event) belongs to a certain topic. Thus, subscribers express
their interest in a particular subject and they receive all the events
published within that particular subject. The most significant restriction
of these systems is the limited selectivity of subscriptions. The
latest systems are called content-based systems. In these systems,
the subscriptions can contain complex queries on event content.

Pub/sub systems try to solve the problem of selective information
dissemination. Recently, there has been a lot of research
on solving the problem of efficiently matching events against
subscriptions. The proposed solutions are either centralized, where
a single broker stores all subscriptions and event matching is done
locally~\cite{aguilera99matching,yfilter,arno} or distributed, where many
brokers need to collaborate to match events with subscriptions because not
all subscriptions are available to every broker~\cite{siena,jedi}. The
latter approach is also referred to as content-based routing because
brokers form a network where events are routed to interested
subscribers based on their content. 

The existing solutions are limited because the matching (routing) is
based on the syntax and not on the semantics of the information
exchanged.  For example, someone interested in buying a car with a
``value of up to 10,000 will not receive notifications about
``vehicles, ``automobiles or even ``cars with ``price of 8,999
because the system has no understanding of the ``price-``value
relationship, nor of the ``car-``automobile-``vehicle
relationship.

In this paper we examine the issues in extending distributed pub/sub
systems to offer semantic capabilities. This is an important aspect to
be studied as components in a pub/sub systems are decoupled and do not
necessary speak the same language.

\section{Local Matching and Content-based Routing}
\label{section::background}
Due to space limitation, we will not provide an extensive background
about pub/sub systems and content-based routing. Instead, we
briefly present the most important concepts that help the reader
understand the ideas conceived in this paper.

The key point in pub/sub systems is that the information sent into the
system by the publisher does not contain the addresses of the
receivers.  The information is forwarded to interested clients based on
the content of the message and clients subscriptions. In a centralized
approach, there is only one broker that stores all subscriptions. Upon
receiving an event, the broker uses a matching algorithm to match the
event against the subscriptions in order to decide which subscribers
want to receive notifications about the event.

Usually, publications are expressed as lists of attribute-value pairs.
The formal representation of a publication is given by the following
expression: \textit{($a_{1}$, $val_{1}$), ($a_{2}$, $val_{2}$), ...,
($a_{n}$, $val_{n}$)}. Subscriptions are expressed as conjunctions of
simple predicates. In a formal description, a simple predicate is
represented as \textit{(attribute\_name relational\_operator value)}. A
predicate \textit{(a rel\_op val)} is matched by an attribute-value
pair \textit{(a, val)} if and only if the attribute names are
identical \textit{(a = a)} and the \textit{(a rel\_op val)} boolean
relation is true. A subscription \textit{s} is matched by a publication
\textit{p}  if and only if all its predicates are matched by some pair
in \textit{p}. In this case we say that the subscription is matched at
syntactic level.

The distributed approach involves a network of brokers that collaborate
in order to route the information in the system based on its content.
In this case, practically, each broker is aware of its neighbours
interests. Upon receiving an event, the broker matches it against its
neighbours subscriptions and sends the event only to the interested
neighbours. Usually, the routing scheme presents two distinct aspects:
subscription forwarding and event forwarding. Subscription forwarding
is used to propagate clients interests in the system, while event
forwarding algorithms decide how to disseminate the events to the
interested clients. Two main optimizations were introduced in the
literature in order to increase the performance of these forwarding
algorithms: subscription covering and advertisements. 

\textbf{Subscription covering}

Given two subscriptions $s_1$ and $s_2$, $s_1$ covers $s_2$ if and only
if all the events that match $s_2$ also match $s_1$. In other words, if
we denote with $E_1$ and $E_2$ the set of events that match subscription
$s_1$ and $s_2$, respectively, then $E_2 \subseteq E_1$.

\begin{table*}[t]
\begin{center}
\begin{tabular}{|p{1.8in}|p{1.7in}|p{1.2in}|}
\hline Subscription \textit{$s_1$} & Subscription \textit{$s_2$} & Covering Relation \\ 
\hline (product = ``computer, brand = ``IBM, price $\leq$ 1600) & (product = ``computer, brand = ``IBM, price $\leq$ 1500) & \textit{$s_1$} covers \textit{$s_2$} \\
\hline (product = ``computer, brand = ``IBM, price $\leq$ 1600) & (product = ``computer, price $\leq$ 1600) & \textit{$s_2$} covers \textit{$s_1$} \\
\hline (product = ``computer, brand = ``IBM, price $\leq$ 1600) & (product = ``computer, brand = ``Dell, price $\leq$ 1500) & \textit{$s_1$} does not cover \textit{$s_2$}, \textit{$s_2$} does not cover \textit{$s_1$}\\
\hline

\end{tabular} \\
\end{center}
\caption{Examples of subscriptions and covering relations}\label{table::covsub}
\end{table*}

If we look at the predicate level, the covering relation can be expressed
as follows: Given two subscriptions $s_1$ = ${p_1}^{1}, {p_2}^{1},
\dots,  {p_n}^{1}$ and $s_2$ = ${p_1}^{2}, {p_2}^{2}, \dots,
{p_m}^{2}$, $s_1$ covers $s_2$ if and only if $\forall {p_k}^{1} \in
s_1,  \exists {p_j}^{2} \in s_2$\footnote{${p_k}^{1}$ and ${p_j}^{2}$ refer to the same attribute} such that if ${p_j}^{2}$ is matched by
some attribute-value pair $(a, val)$, then ${p_k}^{1}$ is also matched by
the same $(a, val)$ attribute-value pair. In other words, $s_2$ has
potentially more predicates and they are more restrictive than those in
$s_1$. Table ~\ref{table::covsub} presents some examples of
subscriptions and the corresponding covering relations.

When a
broker $B$ receives a subscription $s$, it will send it
to its neighbours if and only if it has not previously sent them
another subscription $s$, that covers $s$. Broker
$B$ is ensured to receive all events that match $s$, since it
receives all events that match $s$ and the events that match
$s$ are included in the set of the events that match
$s$. 

\begin{table*}[t]
\begin{center}
\begin{tabular}{|p{1.8in}|p{1.7in}|p{1.2in}|}
\hline Subscription \textit{$s$} & Advertisement \textit{$a$} & Intersection Relation \\ 
\hline (product = ``computer, brand = ``IBM, price $\leq$ 1600) & (product = ``computer, brand = ``IBM, price $\leq$ 1500) & \textit{$a$} intersects \textit{$s$} \\
\hline (product = ``computer,  price $\leq$ 1600) & (product = ``computer, brand = ``IBM, price $\leq$ 1600) & \textit{$a$} intersects \textit{$s$} \\
\hline (product = ``computer, brand = ``IBM, price $\leq$ 1600) & (product = ``computer, brand = ``Dell, price $\leq$ 1500) & \textit{$a$} does not intersect \textit{$s$}\\
\hline

\end{tabular} \\
\end{center}
\caption{Examples of subscriptions, advertisements and intersection relations}\label{table::interads}
\end{table*}

\pagebreak
\textbf{Advertisements} 

Advertisements are used by publishers to announce the set of
publications they are going to publish. Advertisements look exactly
like subscriptions\footnote{However, there is an important distinction
between the predicates in an advertisement and those in a subscription:
the predicate in a subscription are considered to be in a conjunctive
construction, while those in an advertisement behave as in a
disjunctive one.}, but have a different role in the system: they are
used to build the routing path from the publishers to the interested
subscribers.

An advertisement $a$ determines an event $e$ if and only if all
attribute-value pairs match some predicates in the advertisement.
Formally, an advertisement $a$ = ${p_1}^{1}, {p_2}^{1},  \dots, {p_n}^{1}$
determines an event $e$, if and only if $\forall (a,v) \in e,  \exists
p_k \in a$ such that $(a,v)$ matches $p_k$.

An advertisement $a$ intersects a subscription $s$ if and only if the
intersection of the set of the events determined by the advertisement $a$
and the set of the events that match $s$ is a non-empty set. Formally,
at predicate level, an advertisement $a$ = ${a_1}, {a_2}, \dots, {a_n}$
intersects a subscription $s$ = ${s_1}, {s_2}, \dots, {s_n}$ if and only
if $\forall s_k \in s,  \exists a_j \in a$ and some attribute-value pair
$(attr, val)$\footnote{$s_k$ and ${a_j}$ refer to the same attribute
$attr$} such that $(attr, val)$ matches both $s_k$ and $a_j$. Table
~\ref{table::interads} presents some examples of subscriptions and
advertisements and the corresponding intersection relations.

When using advertisements, upon receiving a subscription, each broker
forwards it only to the neighbours that previously sent advertisements
that intersect with the subscription. Thus, the subscriptions are
forwarded only to the brokers that have potentially interesting
publishers.

\section{Towards Semantic-based Routing}
In order to add a semantic dimension to distributed pub/sub systems,
we have to understand how to adapt or map the core concepts and
functionalities of existing solutions for content-based routing to the
new context that involves semantic knowledge.

In this section we first introduce some extensions to the existing
matching algorithms in order to make them semantic-aware and then we
discuss the implications of using such a solution for semantic-based
routing.

\subsection{Semantic Matching}
\label{section::semmatch}
In this section we summarize our approach to make the existing
centralized matching algorithms semantic-aware~\cite{stopss}. Our goal
is to minimize the changes to the existing matching algorithms so that
we can take advantage of their already efficient techniques and to make
the processing of semantic information fast. We describe three
approaches, each adding more extensive semantic capability to the
matching algorithms.  Each of the approaches can be used independently
and for some applications that may be desirable. It is also possible to
use all three approaches together.

The first approach allows a matching algorithm to match events and
subscriptions that use semantically equivalent attributes or
values---synonyms.  The second approach uses additional knowledge about
the relationships (beyond synonyms) between attributes and values to
allow additional matches. More precisely, it  uses a concept hierarchy
that provides two kinds of relations: specialization and
generalization.  The third approach uses mapping functions which allow
definitions of arbitrary relationships between the schema and the
attribute values of the event.

The synonym step involves translating all strings with different names
but with the same meaning to a ``root term. For example, ``car and
``automobile are synonyms for ``vehicle which then
becomes the root term for the three words. This translation is
performed for both subscriptions and events and at both attribute and
value level.  This allows syntactically different events and
subscriptions to match.  This translation is simple and
straightforward. The semantic capability it adds to the
system, although important, may not be sufficient in some situations,
because this approach operates only at attribute and value level
independently and does not consider the semantic relation between
attributes and values.  Moreover, this approach is limited to synonym
relations only.

Taxonomies represent a way of organizing ontological knowledge using
specialization and generalization relationships between different
concepts.  Intuitively, all the terms contained in such a taxonomy can
be represented in a hierarchical structure, where more general terms
are higher up in the hierarchy and are linked to more specialized terms
situated lower in the hierarchy.  This structure is called a ``concept
hierarchy.  Usually, a concept hierarchy contains all terms within a
specific domain, which includes both attributes and values.

Considering the observation that the subscriber should receive only
information that it has precisely requested, we come up with the following
two rules for matching that uses concept hierarchy: (1) the events that
contain more specialized concepts have to match the subscriptions that
contain more generalized terms of the same kind and (2) the events that
contain more generalized terms than those used in the subscriptions do
not match the subscriptions.

In order to better understand these rules, we look at the following
examples. Suppose that we have in the system a subscription:\\
\centerline{$S: (book = StoneAge) AND (subject = reptiles)$.}\\
When the event:\\
\centerline{$E: {(encyclopedia, StoneAge), (subject, crocodiles)}$}\\
is entering the system, it should match the subscription $S$, as the
subscriber asked for more general information that the event provides
(in other words, an $encyclopedia$ is a special kind of $book$ and
$crocodiles$ represent a special kind of $reptiles$). On the other
hand, considering the subscription:\\
\centerline{$S: (encyclopedia=StoneAge) AND (subject=reptiles)$}\\
and the incoming event\\ 
\centerline{$E: {(book, StoneAge), (subject, crocodiles)}$,}\\
the event $E$ should not match the subscription $S$, as the book
contained in the event may be a dictionary or a fiction book (as well as
an encyclopedia). Note that, although the subscription $S$ contains in
its second predicate a value more specialized than that in the event,
the first predicate of the subscription is not matched by the event,
and therefore, the event does not match the subscription. The last rule
prevents an eventual spamming of the subscribers with useless information.

Mapping functions can specify relationships which otherwise cannot be
specified using a concept hierarchy or a synonym relationship. For
example, they can be used to create a mapping between different
ontologies. A mapping function is a many-to-many function that
correlates one or more attribute-value pairs to one or more
semantically related attribute-value pairs.  It is possible to have
many mapping functions for each attribute. We assume that mapping
functions are specified by domain experts. In the future, we are going
to investigate using a fully-fledged inference engine as a more compact
representation of mapping functions and the performance trade off this
entails.

We illustrate the concept of mapping functions with an example. 
Let us say that there is a university
professor X, who is interested in advising new PhD graduate students. In
particular, he is only interested in students who have had 5 or more
years of previous professional experience. Subsequently, he subscribes
to the following:\\
\centerline{$S: (university=Y)AND(degree=PhD)AND(professional\ experience > 4)$}
Specifically, the professor X is looking for students applying to
university Y in the PhD stream with 5 or more years of experience. For
each new student applying to the university, a new event, which contains
(among others) the information about previous work experience, is
published into our system. Thus, an event for a student who had some
work experience would look like\\
\centerline{$E: {(school,Y)(degree,PhD)(work\ experience,true)
(graduation\ date, 1990)}$.}\\
In addition, the system has access to the following mapping function:\\
\centerline{$f_1: (work\ experience,graduation\ date) \rightarrow
professional\ experience$.}\\
You can think of function $f_1$ implemented as a simple difference between
todays date and the date of students graduation and returning that
difference as the value of $professional\ experience$. For the purposes
of the example, $f_1$ assumes that the student has been working since
graduation.  Finally, the result of $f_1$ is appended to event $E$ and
the matching algorithm matches $E$ to professor Xs subscription $S$.

In addition, we can think about events and subscriptions as points in a
multidimensional space~\cite{subspace} where the distance between
points determines a match between an event and a subscription.  This
way it is possible that an event matches a subscription even if some
attribute/value pair of the event is more general than the
corresponding predicate in the subscription as long as the distance
between the event and the subscription, as determined by {\sl all\/}
their constituent attribute-value pairs and predicates respectively, is
within the defined matching range.

To summarize, the synonym stage translates the events and the
subscriptions to a normalized form using the root terms, while the
hierarchy and the mapping stages add new attribute-value pairs to the
events. The new events are matched using existing matching algorithms
against the subscriptions in the system. In conclusion, we say that $e$
semantically matches $s$\footnote{$e$ and $s$ are in their normalized form}
if and only if the hierarchy and the mapping stages can produce an
event $e= e \cup E$\footnote{$E$ represents the set of
attribute-value pairs that are added by the hierarchy and the mapping
stages. Note that $E$ can be an empty set.} that matches $s
$ at syntactic level.

\subsection{Semantic-based Routing}
At first glance, it is apparent that existing algorithms for
subscription and event forwarding can be used with a semantic-aware
matching algorithm in order to achieve semantic-based routing. However,
this approach is not straight forward. In this section we discuss some
open issues that arise from using a semantic-aware matching algorithm
in content-based routing.

\textbf{Subscription covering} 

Although it is defined at syntax level, the covering relation as
presented in Section \ref{section::background} can be used directly
with the semantic matching approach discussed above without any loss
of notifications. In other words, if $s_1$ covers $s_2$ and a certain
broker $B$ will forward only subscription $s_1$ to its neighbours,
it will still receive both events that semantically match $s_1$ and
$s_2$. This happens because the relation between the set of events $E_1$
and $E_2$ that semantically match $s_1$ and $s_2$ respectively, is
preserved, i.e. $E_2 \subseteq E_1$. Truly, if $e$ semantically matches
$s_2$, then the hierarchy and the mapping stages can produce an event
$e$ that matches $s_2$ at syntactic level. If $e$ matches $s_2$ at
syntactic level, then, according to the definition of covering relation,
$e$ matches $s_1$ at syntactic level. Since $e$ is produced by adding
semantic knowledge to $e$, this means that $e$ semantically matches $s_1$,
i.e. $E_2 \subseteq E_1$. Thus, broker $B$ is ensured to receive all
events that semantically match $s_2$, since it receives all events that
semantically match $s_1$ and the events that semantically match $s_2$
are included in the set of the events that semantically match $s_1$.

Although the syntactic covering relation can be used without loss
of notifications, some redundant subscriptions may be forwarded into
the network. This happens because the set of events $E_1$ and $E_2$
that semantically match $s_1$ and $s_2$ can be in the following
relation $E_2 \subseteq E_1$ without necessarily $s_1$ covering $s_2$
at syntax level. In other words, although $s_1$ does not cover $s_2$
at syntactic level, it may cover it semantically speaking. For example,
consider the following subscriptions: $s_1$ = $((product = "printed$
$material") AND (topic = "semantic$ $web"))$ and $s_2$ = $((product =
"book") AND (topic = "semantic$ $web"))$. In this case, all events that
semantically match $s_2$ will also match $s_1$ as a \textit{book} is a
form of \textit{printed material}; thus $E_2 \subseteq E_1$, but $s_1$
does not cover $s_2$ (at syntax level). Therefore, the covering relation
needs to be extended to encapsulate semantic knowledge. One simple way
of transforming the covering relation to be semantic-aware is to use the
hierarchy approach. In this case, subscription $s_1$ will cover $s_2$
as the $printed$ $material$ term is a more general term than $book$.


\textbf{Advertisements}

While the covering relation can be directly used with the semantic
matching algorithms, this is not the case for advertisements. As explained
earlier in this paper, advertisements are used to establish the routing
path from the publishers to the interested subscribers. How the events
are routed in the system depends on the intersection relation between
advertisements and subscription. Consider the following example:
advertisement $a$ = $((product = ``printed material), (price \geq
10))$ and subscription $s$ = $((product = ``book), (price \leq
20))$. Advertisement $a$ does not intersect $s$ at syntactic level because
there is not any predicate $p$ in $a$ and not any attribute-value pair
$(attr, val)$ such that $(attr, val)$ matches both $p$ and the following
predicate $(product = ``book)$ of subscription $s$. (cf. Section
~\ref{section::background}. Thus, the subscription will not be forwarded
towards the publisher that emitted the advertisement. All publications
that will be produced by this publisher will not be forwarded to the
subscriber, although some of them may matched its subscriptions.

\textbf{Distributed semantic knowledge}

The discussion above about subscription covering and advertisements
considered that each broker contains the same semantic knowledge (i.e.
same synonyms, hierarchies and mapping functions). However, the
replication of the same semantic knowledge to all brokers in the system
may not be feasible and it may be detrimental to scalability.

We envision a system where semantic knowledge is distributed between
brokers\footnote{We use the term {\sl broker} and {\sl router}
interchangeably.} in the same way that Internet distributes link
status information using routing protocols. A semantic knowledge
database is equivalent to routing tables in terms of functionality.

\begin{figure}[t]
\centering\includegraphics[width=0.5\textwidth]{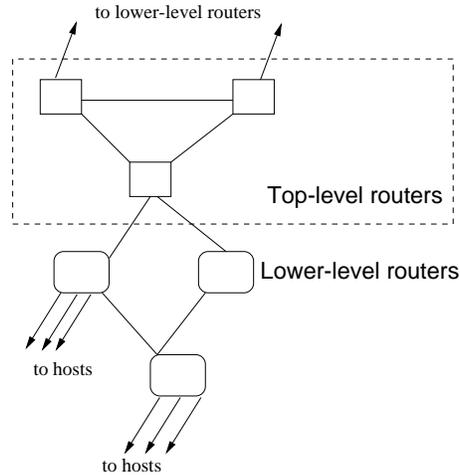}
\caption{Conceptual illustration of a two-level distributed semantic
pub/sub network.  Top-level routers have only high level
descriptions of ontologies from the lower level routers.}
\label{figure::dist-onto}
\end{figure}

The Internet is a hierarchical computer network. At the top of hierarchy
are relatively few routers containing very general information in routing
tables. The tables do not contain information about every host on the
Internet, but only about a few network destinations.  Thus, high
level pre-defined ontological information could be distributed in the
same way among the top routers (Figure~\ref{figure::dist-onto}). It is
difficult to envision what this higher level information will be at this
time, but we only need to take a look at Internet directories such as
Google and Yahoo to get an idea of top level semantic knowledge. Both of
these directories provide a user with only a few key entries as starting
point for exploring the wast Internet information store.  We see top
level brokers exchanging only covering and advertisement information.

Lower in the Internet hierarchy  routers maintain routing tables with
destinations to specific hosts.  Even though top level brokers use a
common ontology, lower level brokers do not have to. For example,
consider two different pairs of communicating applications: financial
and medical.  Financial applications are  exchanging stock quotes,
while medical are exchanging news about new drugs. These two
application use different ontologies. The ontology information for each
application can be distributed between multiple routers.  These low
level brokers will advertise more general descriptions of the
ontologies they have to higher level brokers. Using this information,
any new application will be able to locate the broker with specific
ontologies. Any application wishing to integrate medical and financial
information can create a mapping ontology between the financial and
medical ontologies and provide a general description of the mapping
ontology to higher level broker like in the previous case.  We see that
high level concepts can be used to route information between brokers
who do not have access to specific ontologies. We can look at these
general terms as very terse summaries of ontologies.

Our vision of a large scale semantic-based routing raises many
questions:
\begin{itemize}
\item{\bf top-level routing:} How to bridge multiple distributed ontologies
to enable content routing? How can we avoid or reduce duplication of
ontological information among brokers? What is an appropriate high
level generalization that can bring together different ontologies? How
do semantic routing protocols look like?
\item{\bf lower-level routing:} How to efficiently store ontological
information at routers? Large knowledge databases will probably require
secondary storage beyond what is available at routers. How does this
affect routing? If routers have to use covering at this level how can
they dynamically control the generality of covering to affect network
performance?
\end{itemize}

\section{Related work}
We are not aware of any previous work addressing the semantic routing
problem in pub/sub systems.  Most research on semantic has been done in
the area of heterogeneous database integration. The main problem in this
area is on enabling integration of heterogeneous information systems so
that users can access multiple data sources in an uniform manner. One
way of solving this problem is by using ontologies. Semantic information
systems use an ontology to represent domain-specific knowledge and allow
users to use the ontology terms to construct queries. The query execution
engine accesses the ontology either directly or via an inference engine
in order to optimize the query and generate an execution plan.  Use of an
ontology to generate an execution plan is central in determining the right
source database and method for retrieving the required information. This
allows uniform access to multiple heterogeneous information sources. The
problem of adding semantic capability to pub/sub systems can be seen as an
``inverse problem to the heterogeneous database integration problem.
In semantic pub/sub systems, subscriptions are analogous to queries
and events correspond to data, so now the problem is how to match data
to queries.

Some systems \cite{carnot,oldsemantic} use inference engines
to discover semantic relationships between data from ontology
representations. Inference engines usually have specialized languages
for expressing queries different from the language used to retrieve data,
therefore user queries have to be either expressed in or translated into
the language of the inference engine.  The ontology is either global
(i.e.,~domain independent) or domain-specific (i.e.,~only a single
domain) ontology.  Domain-specific ontologies are smaller and more
commonly found than global ontologies because they are easier to specify.
Additionally, there are systems that use mapping functions exclusively and
do not have inference engines \cite{observer,semval}. In these systems,
mapping functions serve the role of an inference engine.

Web service discovery is a process of matching user needs to provided
services; user needs are analogous to events and provided services
to subscriptions in a pub/sub system.  Web service discovery systems
\cite{sem2,sem3} are functionally similar to a pub/sub system. During
a discovery process,  a web service advertises its capabilities in terms
of its inputs and outputs.  An ontology provides an association between
related inputs or outputs of different web services. A user looks for a
particular web service by searching for appropriate inputs and outputs
according to the users needs. Relevant services are determined by either
exact match of inputs and outputs, or a compatible match according to
ontology relationships.

The main push for using ontologies and semantic information
as means of creating a more sophisticated application
collaboration mechanisms has been from the Semantic Web
community\footnote{www.semanticweb.org}. Recently their focus was on
developing DAML+OIL---a language for expressing, storing and exchange
of ontologies.  Our vision of a distributed semantic pub/sub
system is similar to that of the semantic web.  The issues of distributing
ontological information and bridging of different ontologies are common
to both.

A system for distributed collaboration~\cite{semmcast} creates a
virtual network of proxies (functionally similar to brokers) using IP
multicast connecting both data producers and consumers (users).  Using
common ontology, sources provide descriptions (metadata similar to
subscriptions and events) of multimedia data they are providing and
users provide descriptions of their capabilities.  The metadata is
distributed among proxies to create a {\sl semantic multicast graph}
along which data is distributed to interested users.

To improve scalability, peer-to-peer database systems are looking in the
direction of semantic routing. HyperCuP~\cite{hypercup} uses common
ontology to dynamically cluster peers based on the data they contain.
A cluster is identified using a more general concept then those
associated with its members in the ontology. Concepts in the ontology
map to cluster addresses so a node can determine appropriate route for
a query by looking up more general concepts of the query terms in the
concept hierarchy.  Edutella~\cite{edutella} uses query hubs
(functionally similar to brokers) to collect user metadata and present
the peer-to-peer network as a virtual database which users query. All
queries are routed though a query hub which forwards queries only to
those nodes that can answer it.

\section{Conclusions}
In this paper we underline the limits of matching and content-based
routing at syntactic level in pub/sub systems. We propose
a solution for achieving semantic capabilities for local matching and
look into the implications of using such a solution for content-based
routing. We also present our vision on next-generation semantic-based
routing. Our intent was to give rise to questions and ideas in order
to improve existing content-based routing approaches and make them
semantic-aware.

{\small
\bibliographystyle{unsrt}
\bibliography{srt}
}

\end{document}